\documentclass[12pt,a4paper,final]{iopart}
\usepackage{iopams}
\expandafter\let\csname equation*\endcsname\relax
\expandafter\let\csname endequation*\endcsname\relax

\usepackage{graphicx}
\usepackage{amsmath}
\usepackage{amssymb}
\usepackage{bbm}
\usepackage{bm}

\usepackage{xspace}

\newcommand{\ket}[1]{| #1 \rangle}

\newcommand{\rb}[1]{\left( #1 \right)}

\newcommand{\eq}[1]{Eq.~(\ref{#1})}
\newcommand{\fig}[1]{Fig.~\ref{#1}}

\graphicspath{{figures/}}

\newcommand{\PT}{$\mathcal{PT}$\xspace}

\newcommand{\Rb}{$^{87}$Rb\xspace}

\begin{document}
\title{Perpetual emulation threshold of \PT-symmetric Hamiltonians}

\author{D. Trypogeorgos, A. Vald\'es-Curiel, I. B. Spielman}
\address{Joint Quantum Institute, University of Maryland and National
Institute of Standards and Technology, College Park, Maryland, 20742, USA}

\author{C. Emary}
\address{
Joint Quantum Centre Durham-Newcastle, School of Mathematics, Statistics and Physics, Newcastle University, Newcastle upon Tyne NE1 7RU, UK
}

\vspace{10pt}
\begin{indented}
\item[]\today
\end{indented}

\begin{abstract}
We describe a technique to emulate a two-level \PT-symmetric spin Hamiltonian, replete with gain and loss, using only the unitary dynamics of a larger quantum system.
This we achieve by embedding the two-level system in question in a subspace of a four-level Hamiltonian.
Using an \textit{amplitude recycling} scheme that couples the levels exterior to the \PT-symmetric subspace, we show that it is possible to emulate the desired behaviour of the \PT-symmetric Hamiltonian without depleting the exterior, reservoir levels.
We are thus able to extend the emulation time indefinitely, despite the non-unitary \PT dynamics.
We propose a realistic experimental implementation using dynamically decoupled magnetic sublevels of ultracold atoms.
\end{abstract}

\maketitle
%%%%%%%%%%%%%%%%%%%%%%%%%%%%%%%%%%%%%%%%%%%%%%%%%%%%%%%%%%%%%%%%%%%%%%%%%

\section{Introduction}
Quantum theory is our most successful description of nature.
It describes the dynamics of closed systems by means of Hermitian Hamiltonians.
The Hermitian constraint leads to unitary evolution and a real, and hence measurable, eigenvalue spectrum.
Remarkably, this constraint is stronger than necessary, and a subclass of non-Hermitian Hamiltonians also affords real-valued eigenvalues as long as the Hamiltonian operator commutes with the joint parity-time \PT operator.
These models emerged from perturbative approaches in quantum field theory~\cite{bender_novel_1988,bender_logarithmic_1987} and can have all-real eigenvalues in some parameter regimes~\cite{dorey_ode/im_2007,dorey_spectral_2001}.
In such models, a free parameter of the Hamiltonian drives a phase transition between two regimes where the \PT symmetry is either \textit{unbroken} or \textit{broken}, leading to real and complex eigenvalues respectively~\cite{bender_making_2007}.

The language of \PT-symmetric Hamiltonians can offer a simplified, alternative formulation of the Lindblad equation description of open systems~\cite{dast_quantum_2014}, and was used to study numerous interesting and disparate phenomena ranging from photosynthesis~\cite{eleuch_gain_2017-1} to the slowing down of decoherence in the vicinity of the critical point~\cite{gardas_$mathcalpt$-symmetric_2016}.
Generally, it is possible to derive a non-Hermitian Hamiltonian from a given Lindblad equation, that is a valid description of the systems' dynamics until the probability of a quantum jump becomes too large.

A number of experiments in the classical domain have demonstrated \PT-symmetric systems, e.g., using coupled optical waveguides~\cite{peng_paritytime-symmetric_2014,ruter_observation_2010,guo_observation_2009}, and Floquet systems~\cite{chitsazi_experimental_2017-1}.
The first demonstration in the quantum domain used Sinai billiards~\cite{gao_observation_2015,cao_dielectric_2015}.
The physics of these systems becomes even richer close to the critical point between the broken and unbroken regime; the phase associated with encircling the critical point is similar to the Berry phase associated with a Dirac point in non-trivial topological materials~\cite{klett_relation_2017-1,menke_topological_2017-1}.

In a series of papers~\cite{Kreibich2013,Kreibich2014,Gutoehrlein2015}, Wunner and coworkers showed that it is possible in principle to reproduce the dynamics of a two-level \PT-symmetric Hamiltonian with a four-level Hermitian system.
In this configuration two of the levels map to the \PT-symmetric levels and the remaining two act as probability amplitude source and sink.
The embedded \PT-subspace is coupled to the rest of the system using time-dependent transition matrix elements that emulate the non-unitary particle flow between the subspaces; we refer to this scheme as coherent emulation.
Here we extend these works, initially by adopting a different calculational approach that permits analytic closed-form solutions in these four-level systems.
This approach makes it explicit that the emulation of the two-level \PT-symmetric dynamics in these works is always limited in time, and that it breaks down due to a depletion of the probability amplitude in the source level.

We then consider an extended scheme that includes an additional coupling between source and sink levels, to counter the depletion of the source level.
We show that this amplitude recycling increases the emulation time for any given set of parameters.
Moreover, when the coupling strength of the amplitude-recycling field exceeds a critical value, the duration of the emulation is extended indefinitely.
Above this threshold the system dynamics dramatically change such that a genuinely periodic behaviour occurs.
Implementation of this scheme requires the solution of a pair of coupled, non-linear differential equations to pre-compute the time-dependent Hamiltonian required to emulate \PT-symmetric behaviour in the target subspace.
Our protocol therefore consists of a classical computation device to design a control sequence, different for all initial conditions, and a quantum system to implement the sequence.
We use these results as the basis for the discussion of a realisation of the \PT-symmetric dynamics in pseudospins corresponding to internal states of the groundstate manifold of alkali atoms cooled to degeneracy.

%%%%%%%%%%%%%%%%%%%%%%%%%%%%%%%%%%%%%%%%%%%%%%%%%%%%%%%%%%%%%%%%%%%%%%%
%%%%%%%%%%%%%%%%%%%%      2LS \PT Ham     %%%%%%%%%%%%%%%%%%%%%%%%%%%%%%
%%%%%%%%%%%%%%%%%%%%%%%%%%%%%%%%%%%%%%%%%%%%%%%%%%%%%%%%%%%%%%%%%%%%%%%
\section{Two-level \PT-symmetric Hamiltonian \label{SEC:PT}}
We focus on an extensively studied minimal model~\cite{bender_must_2003,bender_complex_2002,Wang2013,deffner_jarzynski_2015,ruter_observation_2010,gao_observation_2015}, that captures all the relevant information of such systems,
\begin{equation}\label{eq:hpt}
  h =
  \rb{
    \begin{array}{cc}
      -i \Gamma & \Lambda \\
      \Lambda & i \Gamma
    \end{array}
  }
  ,
\end{equation}
and aim to describe the full dynamics governed by $h$ for all times and $\Gamma,\,\Lambda>0$.
The parity operator for this model is $\mathcal{P}=\sigma_x$, where $\sigma_i$, $i=x,y,z$ are the Pauli operators, and time reversal gives $\mathcal{T} h \mathcal{T}^{-1} =  h^*$.
For this Hamiltonian, \PT-symmetry implies
$
  \mathcal{P} h -  h^*\mathcal{P} = 0
$, which is manifestly obeyed by \eq{eq:hpt}.
The eigenvalues of $h$ are
$
  \epsilon_\pm = \pm  \alpha/2
$,
where
$
  \alpha = 2 \sqrt{\Lambda^2 - \Gamma^2}
$
is the level splitting.
This sets a natural energy scale for the system and allows us to use dimensionless time $\tau=\alpha t$; coupling $\lambda=2\Lambda/\alpha$; and gain $\gamma=2\Gamma/\alpha$.
We assume here that we stay in the \PT-unbroken phase such that $\alpha$ is real, i.e. $ \Gamma < \Lambda$.
As with Hermitian Hamiltonians we define a time evolution operator $U(\tau)$ that propagates the state $\psi(\tau) = (\psi_1(\tau), \psi_2(\tau))^T$ to time $\tau$:
\begin{equation}
  U(\tau)
  =
   \cos\rb{\tau/2} \mathbb{I}
   -
   \sin\rb{\tau/2}
     \rb{\gamma \sigma_z + i \lambda \sigma_x}
     ,
\end{equation}
where $\mathbb{I}$ is the $2\times2$ unit matrix.
We consider initial conditions with real-valued amplitudes throughout $\psi(0) = \rb{\cos (\theta/2), \sin (\theta/2)}$.

\begin{figure}[tb]
  \begin{center}
  \includegraphics[clip=true]{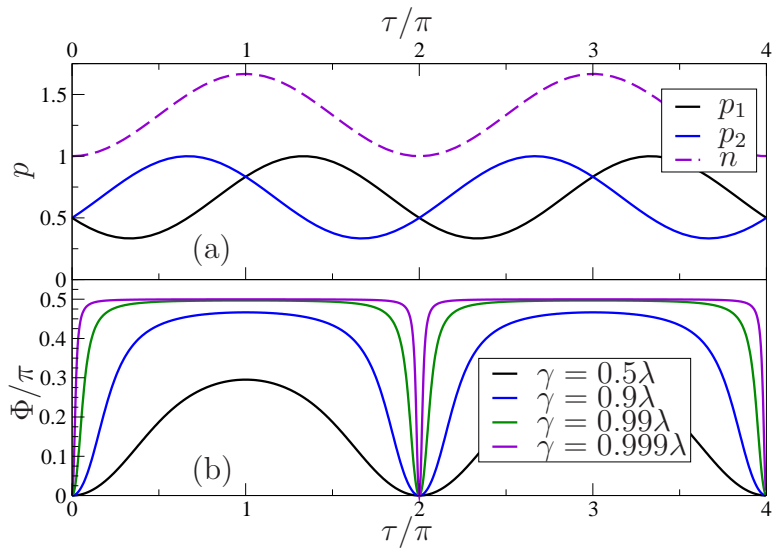}
   \caption{
     (a) The two populations $p_1=|\psi_1|^2$ (black) and $p_2=|\psi_2|^2$ (blue) of the \PT-symmetic system with  $\psi_1(0)= \psi_2(0) =1/\sqrt{2}$ and $\gamma = 0.5 \lambda$.
     The norm $n$ from \eq{EQ:nspesh} is also shown.
     (b) The phase difference $\Phi$ of \eq{EQ:phispesh} as a function of time for various $\gamma$.
     The time axis here is scaled with \PT level-splitting $\alpha$.
     Relative to this scale, the minima of the phase become ever sharper as $\gamma$ increases.
     \label{FIG:phasediff}
   }
 \end{center}
\end{figure}

Several aspects of this solution are of particular interest here.
First, we consider the norm
$
  n(\tau) \equiv |\psi_1(\tau)|^2 + |\psi_2(\tau)|^2
$
and the atomic magnetisation
$
  w(\tau) \equiv |\psi_1(\tau)|^2 - |\psi_2(\tau)|^2
$.
Although our technique is valid for all initial states, for brevity of exposition we base our subsequent discussion on the symmetric case when $\theta = \pi/2$.
In this case, we obtain
\begin{align}
   n(\tau) &= \lambda^2 - \gamma^2\cos(\tau)
  ;
  \label{EQ:nspesh}
  \\
  w(\tau) &= -
  \gamma \sin(\tau)
  .
  \label{EQ:Wspesh}
\end{align}
For $\gamma \to 0$, the norm becomes $\lambda^2 \to 1$, and the magnetisation disappears.
In this limit, there is no evolution since the initial state is an eigenstate of $h$.
For $\theta=\pi/2$, oscillations in the magnetisation only occur when $h$ is non-Hermitian, and the oscillation frequency is given by the splitting $\alpha$.
Also of interest is the relative phase between wave-function components which, again for $\theta=\pi/2$, is defined by
\begin{equation}
  \tan{(\Phi(\tau))} =
      \gamma \lambda(1 - \cos (\tau))
  \label{EQ:phispesh}
  .
\end{equation}
This phase difference is plotted in \fig{FIG:phasediff}.

%%%%%%%%%%%%%%%%%%%%%%%%%%%%%%%%%%%%%%%%%%%%%%%%%%%%%%%
%%%%%%%%%%%%%%%%%%%%%%%%%%%%%%%%%%%%%%%%%%%%%%%%%%%%%%%

Two important timescales emerge from these equations.
The first is the period of the \PT-symmetric oscillations in \eq{EQ:Wspesh}, which, with our choice of scaling for $\tau$, is simply $2\pi$.
The second corresponds to important changes that take place at shorter times.
These are apparent in the phase difference in \eq{EQ:phispesh}, plotted in \fig{FIG:phasediff}.
For small $\gamma/\lambda$, we may approximate
$
  \Phi(\tau) \approx
  2 \gamma/ \lambda \sin^2 (\tau/2)
$
and the dynamics unfold with the same period as the magnetisation of \eq{EQ:Wspesh}.
However, for increasing $\gamma$, the nonlinearity of the arctan function becomes important, and gives rise to sharp changes in the relative phase.
Expanding the phase about one of its minima, we find
$
  \Phi(\tau) \sim \lambda \gamma (\tau-\tau_\mathrm{min})^2/2
$.
Thus the characteristic timescale associated with these minima is
$
 \tau_\mathrm{sharp} = (\lambda \gamma/2)^{-1/2} \sim \sqrt{2}/ \gamma \ll 1
$
in the $\gamma\to\lambda$ limit.
These two contrasting timescales set strict requirements for any experimental implementation of this scheme.

%%%%%%%%%%%%%%%%%%%%%%%%%%%%%%%%%%%%%%%%%%%%%%%%%%%%%%%%%%%%%%%%%%%%%%%
%%%%%%%%%%%%%%%%%%%%      EMULATION      %%%%%%%%%%%%%%%%%%%%%%%%%%%%%%
%%%%%%%%%%%%%%%%%%%%%%%%%%%%%%%%%%%%%%%%%%%%%%%%%%%%%%%%%%%%%%%%%%%%%%%
\section{Emulation with a four-level system \label{SEC:4LS}}

Given that our understanding of Nature is in terms of Hermitian dynamics, we seek to create designer subspaces of larger systems that evolve according to a desired \PT-symmetric Hamiltonian.
We simulate the non-unitary dynamics of \eq{eq:hpt} using a four-level system with wavefunction $\phi(\tau) = \left(\phi_0(\tau),\phi_1(\tau),\phi_2(\tau),\phi_3(\tau)\right)^T$.
We encode the \PT dynamics in the two central levels $\phi_1$, and $\phi_2$, and use $\phi_0$ and  $\phi_3$ as the sink and source levels.
Our `emulator' will thus be a system with dimensionless Hamiltonian \footnote{The corresponding physical Hamiltonian is $H'(t) = \frac{1}{2} \alpha H(\alpha t)$.}
\begin{equation}
  H(\tau) =
  \rb{
    \begin{array}{cccc}
      \delta_0(\tau) & \Omega_{01}(\tau) & 0 & \Omega_{03}  \\
      \Omega_{01}(\tau) & 0 & \lambda & 0 \\
      0 & \lambda & 0 & \Omega_{23}(\tau) \\
      \Omega_{03} & 0 &  \Omega_{23}(\tau) & \delta_3(\tau)
    \end{array}
  }
  .
\end{equation}
These couplings are sufficient to completely emulate the time-dynamics of \PT-symmetric Hamiltonians as in \eq{eq:hpt}.
The dimensionless detunings $\delta_0(\tau)$, $\delta_3(\tau)$ and the couplings $\Omega_{01}(\tau)$, $\Omega_{23}(\tau)$ are time-dependent functions that we choose such that the behaviour of the two levels in the central subspace matches that of the \PT-symmetric system, i.e. such that $\phi_1(\tau)  = \psi_1(\tau)$ and $\phi_2(\tau)  = \psi_2(\tau)$.
The static coupling $\Omega_{03}$ between the source level $\phi_3$ and sink level $\phi_0$ replenishes the source population by transferring particles from the sink level.
We initialise the system such that the population of the central subspace always starts as $\psi_1(0)^2 + \psi_2(0)^2 = 1$; this implies a rescaling of $\phi_1,\,\phi_2$ by an appropriate factor so that it is normalised to unity at $t=0$.

We determine the dynamic detunings and couplings using the following procedure.
We first split the wavefunction components into real and imaginary parts, $\phi_i = \phi_i^\mathrm{R} + i \phi_i^\mathrm{I}$, and use the Schr\"odinger equation for our four-level system to get the equations:
\begin{equation}
\begin{aligned}
\left(\dot{\phi}_0^R, \dot{\psi}_1^R, \dot{\psi}_2^R, \dot{\phi}_3^R\right)^T  &= H(\tau)\left(\phi_0^I, \psi_1^I, \psi_2^I, \phi_3^I\right)^T; \\
\left(\dot{\phi}_0^I, \dot{\psi}_1^I, \dot{\psi}_2^I, \dot{\phi}_3^I\right)^T  &= - H(\tau)\left(\phi_0^R, \psi_1^R, \psi_2^R, \phi_3^R\right)^T,
\end{aligned}
\label{EQ:EOM}
\end{equation}
where we have replaced $\phi_1$ and $\phi_2$ with their target wavefunctions $\psi_1$ and $\psi_2$, and where $\dot\phi_i = d\phi_i/d\tau$.
We then write the second derivatives of $\psi(\tau)$ with respect to time,
\begin{equation}
\begin{aligned}
  \ddot{\psi}_1^\mathrm{R} - \dot{\Omega}_{01}(\tau) \phi_0^\mathrm{I}
    - \Omega_{01}(\tau) \dot{\phi}_0^\mathrm{I} - \lambda \dot{\psi}_2^\mathrm{I} &= 0; \\
  \ddot{\psi}_2^\mathrm{R} - \lambda \dot{\psi}_1^\mathrm{I}
    - \dot{\Omega}_{23}(\tau) \phi_3^\mathrm{I} - \Omega_{23}(\tau) \dot{\phi}_3^\mathrm{I} &= 0,
\end{aligned}
\label{EQ:EOMd3}
\end{equation}
for the real part and,
\begin{equation}
\begin{aligned}
  \ddot{\psi}_1^\mathrm{I} + \dot{\Omega}_{01}(\tau) \phi_0^\mathrm{R}
    + \Omega_{01}(\tau) \dot{\phi}_0^\mathrm{R} + \lambda \dot{\psi}_2^\mathrm{R} &= 0; \\
  \ddot{\psi}_2^\mathrm{I} + \lambda \dot{\psi}_1^\mathrm{R}
    + \dot{\Omega}_{23}(\tau) \phi_3^\mathrm{R} + \Omega_{23}(\tau) \dot{\phi}_3^\mathrm{R} &= 0,
\end{aligned}
\label{EQ:EOMd7}
\end{equation}
the imaginary part respectively.
We use the equations for $\dot\psi_{1,2}$ in Eqs.~(\ref{EQ:EOM}), together with Eqs.~(\ref{EQ:EOMd3}), (\ref{EQ:EOMd7}) to solve for $\phi_{0,3}$, and $\dot\phi_{0,3}$, which we then eliminate from the equations for $\dot\phi_{0,3}$ in Eqs.~(\ref{EQ:EOM}).
From this set of four equations we obtain explicit expressions for $\delta_0(\tau)$ and $\delta_3(\tau)$ in terms of $\Omega_{01}(\tau)$, $\Omega_{23}(\tau)$ plus known quantities, together with a pair of coupled first-order differential equations
\begin{equation}
\begin{aligned}
  \dot{\Omega}_{01} &= f_1 \Omega_{01} + f_2 \Omega_{01}^3 + f_3 \Omega_{03} \Omega^{-1}_{23} \Omega_{01}^2
  ;
  \\
  \dot{\Omega}_{23} &= g_1 \Omega_{23} + g_2 \Omega_{23}^3 + g_3 \Omega_{03} \Omega^{-1}_{01} \Omega_{23}^2
  ,
\end{aligned}
\label{EQ:JJeqns}
\end{equation}
where $f_i\equiv f_i(\tau)$ and $g_i\equiv g_i(\tau)$ are functions defined solely in terms of $\phi_1$, $\phi_2$ and their derivatives.

Our general solution procedure starts therefore by assuming initial values of the couplings and solving the differential equations for  $\Omega_{01}(\tau)$ and $\Omega_{23}(\tau)$.
From these, we obtain $\delta_0(\tau)$ and $\delta_3(\tau)$.
Finally, we determine initial values of the source and sink levels, $\phi_3(\tau=0)$  and  $\phi_0(\tau=0)$, by solving the equations for $\dot\psi_{1,2}$ at $t=0$.
For simplicity in the following we will assume the initial couplings to be equal: $\Omega_{01}(\tau=0) = \Omega_{23}(\tau=0) = \Omega_\mathrm{init}$.

%%%%%%%%%%%%%%%%%%%%%%%%%%%%%%%%%%%%%%%%%%%%%%%%%%%%%%%%%%%%%%%%%%%%%%%
%%%%%%%%%%%%%%%%%%%%      NO REPUMP      %%%%%%%%%%%%%%%%%%%%%%%%%%%%%%
%%%%%%%%%%%%%%%%%%%%%%%%%%%%%%%%%%%%%%%%%%%%%%%%%%%%%%%%%%%%%%%%%%%%%%%
\section{No amplitude recycling \label{SEC:norepump}}

%%%%%%%%%%%%%%%%%%%%%%%%%%%%%%%%%%%%%%%%%%%%%%%
\begin{figure}[tb]
  \begin{center}
  \includegraphics[clip=true]{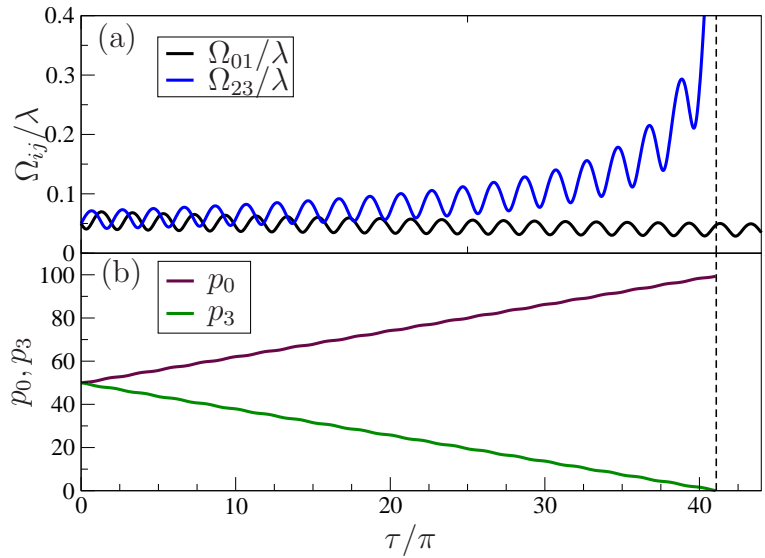}
   \caption{
     (a)
     The two time-dependent couplings $\Omega_{01}(\tau)$ (black) and $\Omega_{23}(\tau)$ (blue) required for emulation, as a function of dimensionless time $\tau/\pi$ for parameters $\gamma = 0.5 \lambda$, $\Omega_{01}(0)=\Omega_{23}(0)=\Omega_\mathrm{init} = 0.05\lambda$, $\phi_1(0)=\phi_2(0)=1/\sqrt{2}$
     and no amplitude recycling  $\Omega_{03}=0$.
     (b)
     The corresponding sink and source populations $p_0=|\phi_0|^2$ (maroon) and $p_3=|\phi_3|^2$ (green).
     At time $\tau = 41.09 \pi$ (indicated with the vertical dashed line), the coupling $\Omega_{23}$ diverges as the source level is depleted.
     This breakdown is a common feature of all solutions without amplitude recycling.
     \label{FIG:eg_no_repump}
   }
 \end{center}
\end{figure}
%%%%%%%%%%%%%%%%%%%%%%%%%%%%%%%%%%%%%%%%%%%%%%%

While Eqs.~(\ref{EQ:JJeqns}) must in general be solved numerically, in the absence of amplitude recycling, $\Omega_{03}=0$, they decouple and admit analytic solutions.
Here we focus on initial conditions with equal amplitude in each of the target PT symmetric states, giving
\begin{equation}
  \rb{\frac{\Omega_{23}(\tau)}{\Omega_\mathrm{init}}}^2
  =
    \frac{
      \lambda^2 -\gamma^2\cos \tau + \gamma \sin \tau
    }{
       1 - \Omega_\mathrm{init}^2
        \left(
          1 + \lambda^2\tau/\gamma - \cos \tau - \gamma \sin \tau
        \right)
    },
  \label{EQ:J23full}
\end{equation}
and
\begin{equation}
    \delta_{0,3}(\tau)
    =
    \frac{
      \lambda
    }
    {\lambda^2
        - \gamma^2 \cos \tau
        \mp\gamma \sin \tau
    }
    \label{EQ:E0}
    ,
\end{equation}
where $\Omega_{01}(\tau) = \Omega_{23}(-\tau)$ (so that when time is reversed the source acts as a sink and vice versa), and $\delta_0(\tau)$ takes the upper sign and $\delta_3(\tau)$ the lower.
The dynamic couplings are plotted for representative parameters in \fig{FIG:eg_no_repump}a, making clear that the coupling $\Omega_{23}$ diverges.
Physically, this break down arises due to the depletion of source level, as shown in \fig{FIG:eg_no_repump}b.
As the population of this level decreases, a stronger coupling $\Omega_{23}$ is required to maintain the constant probability flux at rate $\gamma$ into the system, and this necessarily stops when the level is depleted.
%

%%%%%%%%%%%%%%%%%%%%%%%%%%%%%%%%%%%%%%%%%%%%%%%
\begin{figure}[tb]
  \begin{center}
   \includegraphics[clip=true]{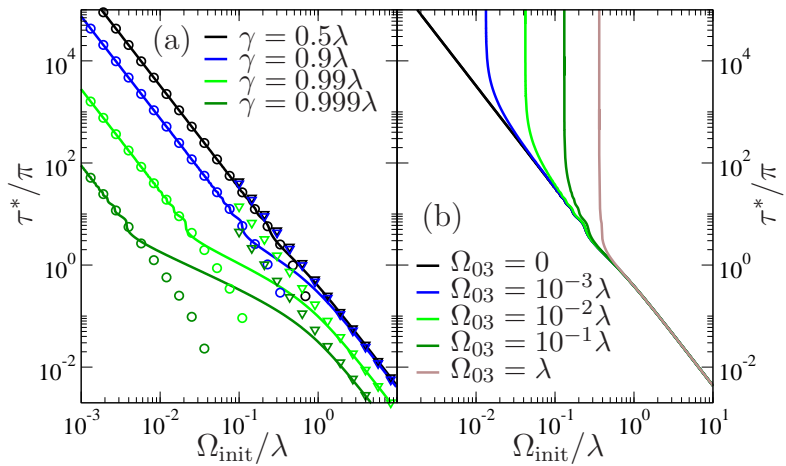}
   \caption{
     The breakdown time $\tau^*$ as a function of the initial coupling strength $\Omega_\mathrm{init}$.
     (a) Case without amplitude recycling, $\Omega_{03}=0$, for several values of $\gamma/\lambda$.
     The solid lines represent the exact solutions; circles, the approximation of \eq{EQ:tsing}; and triangles,  the approximation of \eq{EQ:tsing2}.
     Lowering the coupling strength $\Omega_\mathrm{int}$ extends the duration of the emulation.
     (b) The breakdown time with amplitude recycling for a range of amplitude recycling strengths $\Omega_{03}$ and for fixed $\gamma=0.5\lambda$.
     For large initial coupling strength $\Omega_\mathrm{init}$, the amplitude recycling field has little effect.
     As $\Omega_\mathrm{init}$ is lowered, however, we see a dramatic divergence of the breakdown time which corresponds to the onset of periodic behaviour.
     Parameters not explicitly mentioned are the same as in \fig{FIG:eg_no_repump}
     \label{FIG:tsing_no_repump}
   }
 \end{center}
\end{figure}
%%%%%%%%%%%%%%%%%%%%%%%%%%%%%%%%%%%%%%%%%%%%%%%

Let us denote as $\tau^*$ the time at which $\Omega_{23}$ diverges, the maximum time we can expect our emulation to run.
We obtain $\tau^*$ as the smallest time at which the denominator in \eq{EQ:J23full} vanishes, and plot it in \fig{FIG:tsing_no_repump}a as a function of the starting coupling $\Omega_\mathrm{init}$ for different values of $\gamma$.
For $\gamma$ in the range $0.2\lambda \lesssim \gamma \lesssim 0.8\lambda$, changing $\gamma$ does not alter the $\tau^*$ very much.
However, increasing $\gamma$ above $0.8\lambda$ leads to a significant drop in maximum emulation time; we observe similar behaviour for decreasing $\gamma \lesssim 0.2$.
We can obtain a simple approximation to the breakdown time by setting the oscillating terms in the denominator of \eq{EQ:J23full} to zero.
This gives
\begin{equation}
  \tau^* \approx
  \frac{
    \gamma\rb{1 - \Omega_\mathrm{init}^2}
  }{
    (\Omega_\mathrm{init}\lambda)^2
  }
  \label{EQ:tsing}
  ,
\end{equation}
valid for $\Omega_\mathrm{init} \ll 1$.
In the opposite regime, $\Omega_\mathrm{init}\gg 1$, we obtain an alternative approximation
\begin{equation}
  \tau^* \approx
    \gamma
  /
    \Omega_\mathrm{init}^2
  \label{EQ:tsing2}
  ,
\end{equation}
by expanding the denominator to first order in time, and setting the result equal to zero.
Both these expressions are good approximations in their respective regions of validity as shown in \fig{FIG:tsing_no_repump}a.

From the limits of Eqs.~(\ref{EQ:tsing}), (\ref{EQ:tsing2}), as well as from the full results in \fig{FIG:tsing_no_repump}a, we see that the dominant behaviour of the breakdown time $\tau^*$ is an approximate scaling with $\Omega_\mathrm{init}^{-2}$.
Thus, in principle, we can always arrange our initial coupling strength to enable the emulation to cover any time interval of interest.
However, making $\Omega_\mathrm{init}$ arbitrarily small is not without cost: the initial values of the source and sink populations required by the emulation are
\begin{equation}
   \phi_0(0) = \frac{-i \gamma \phi^\mathrm{I}_1(0)}{\Omega_\mathrm{init}}
   \quad
   \mathrm{and}
   \quad
   \phi_3(0) = \frac{i \gamma \phi^\mathrm{I}_2(0)}{\Omega_\mathrm{init}}
   \label{EQ:initialphi}
   .
\end{equation}

Since the total probability is not contained in the \PT-symmetric subspace, we define the \PT-symmetric fraction in terms of the populations $p_i=|\phi_i|^2$ as
\begin{equation}
  r \equiv \underset{\tau}{\mathrm{min}} \left[ \frac{p_1(\tau)+p_2(\tau)}{\sum_i p_i(\tau)}\right]
  \label{EQ:sigrat}
  .
\end{equation}
This expresses the minimum probability of finding the system in the \PT-symmetric subspace.
This definition is motivated by an experimental resource limit since detecting the probability distribution in all states becomes increasingly cumbersome as \PT-symmetric fraction gets smaller.
The \PT-symmetric fraction is equal to $r =\underset{\tau}{\mathrm{min}}\,\, n(\tau)/N_0$, where $n(\tau)$ is the norm of \eq{EQ:nspesh} and $N_0 = \sum_i p_i(0)$ is the total population at time $\tau=0$, which is a conserved quantity.
Without amplitude recycling and with symmetric initial conditions,  \eq{EQ:nspesh} and \eq{EQ:initialphi} imply
\begin{equation}
  r =
  \underset{\tau}{\mathrm{min}}
  \left[
  \frac{
    \lambda^2 - \gamma^2\cos(\tau)
  }{
    1 + (\gamma/\Omega_\mathrm{init})^2
  }
  \right]
  =
   \frac{
    1
  }{
   1 + (\gamma/\Omega_\mathrm{init})^2
  }
  \label{EQ:rmin}
  .
\end{equation}
In the limit $\Omega_\mathrm{init}/\gamma \ll 1$, this gives $r \sim \Omega_\mathrm{init}^2 / \gamma^2 $.
Thus, decreasing $\Omega_\mathrm{init}$ to extend the emulation time necessarily leads to a corresponding decrease in the \PT-symmetric fraction.

%%%%%%%%%%%%%%%%%%%%%%%%%%%%%%%%%%%%%%%%%%%%%%%%%%%%%%%%%%%%%%%%%%%%%%%
%%%%%%%%%%%%%%%%%%%%      w/ REPUMP      %%%%%%%%%%%%%%%%%%%%%%%%%%%%%%
%%%%%%%%%%%%%%%%%%%%%%%%%%%%%%%%%%%%%%%%%%%%%%%%%%%%%%%%%%%%%%%%%%%%%%%
\section{Amplitude recycling and the perpetual emulation threshold \label{SEC:repump}}

We now consider the effects of the amplitude recycling coupling field.
In this case, Eqs.~(\ref{EQ:JJeqns}) for $\Omega_{01}$ and $\Omega_{23}$ do not decouple and we obtain our results through numerical integration.

Figure~\ref{FIG:tsing_no_repump}b shows the breakdown time $\tau^*$ of the emulation as a function of the initial coupling $\Omega_\mathrm{init}$ for several values of the amplitude recycling strength.
For large $\Omega_\mathrm{init}$, the amplitude recycling does not significantly affect the breakdown time.
However, as $\Omega_\mathrm{init}$ decreases, the breakdown time increases until it rapidly diverges at a value that increases with the amplitude-recycling strength.
%%%%%%%%%%%%%%%%%%%%%%%%%%%%%%%%%%%%%%%%%%%%%%%
\begin{figure}[tb]
  \begin{center}
  \includegraphics[clip=true]{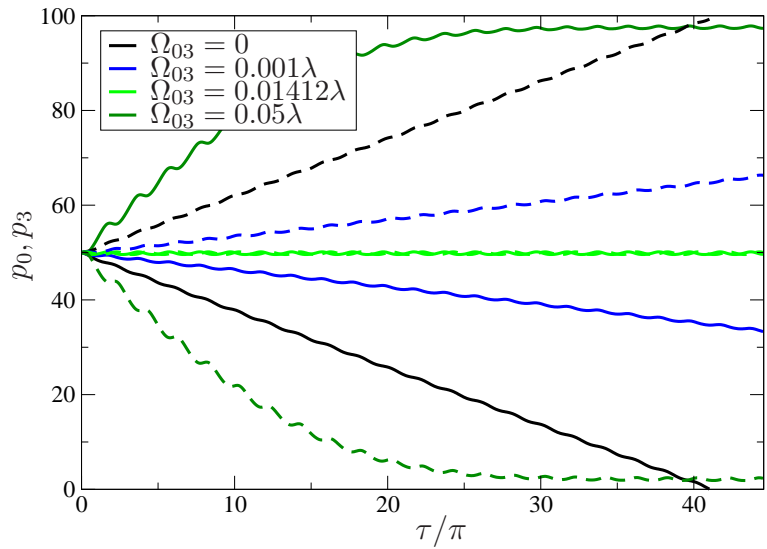}
  \caption{
    Populations $p_0=|\phi_0|^2$ (dashed) and $p_3=|\phi_3|^2$ (solid) as a function of time for different values of $0\le \Omega_{03} \le 0.05\lambda$ and other parameters as in \fig{FIG:eg_no_repump}.
    Increasing $\Omega_{03}$ decreases the overall gradient of $p_3$ until at the critical value $\Omega_{03}=0.01412\lambda$ the cycle-averaged values of $\Omega_{01}$ and $\Omega_{23}$ are constant and equal to one another.  In this case we have purely periodic motion without decay.
    With further increase in $\Omega_{03}$, $p_3$ becomes larger than $p_0$ and in the long-time limit, we observe sustained oscillations of the \PT-symmetric subspace with no breakdown of the emulation.
     \label{FIG:pops_with_repump}
   }
 \end{center}
\end{figure}
%%%%%%%%%%%%%%%%%%%%%%%%%%%%%%%%%%%%%%%%%%%%%%%

Figure~\ref{FIG:pops_with_repump} shows the populations of the sink and source levels as a function of time for several values of the amplitude recycling strength $\Omega_{03}$ and for fixed initial coupling $\Omega_\mathrm{init} = 0.05\lambda$.
As $\Omega_{03}$ is initially increased from zero, the decline of $p_3$ with time becomes shallower.
The solution nevertheless terminates with $p_3=0$, as in the $\Omega_{03}=0$ case.
This trend continues until the critical value $\Omega_{03}= 0.01412\lambda$ of the amplitude recycling strength is reached and the cycle-averaged gradients of $p_0$ and $p_3$ are both zero.
At this point, the behaviour of the populations is purely periodic; they never drop to zero and the coupling $\Omega_{23}$ does not diverge.
For yet larger values of $\Omega_{03}$ the populations reverse order such that $p_3>p_0$, and, in the long-time limit, become periodic once again.
The emulated \PT-symmetric dynamics are correct for all times when $\Omega_{03}$ exceeds its critical value.

\begin{figure}[tb]
  \begin{center}
  \includegraphics[clip=true]{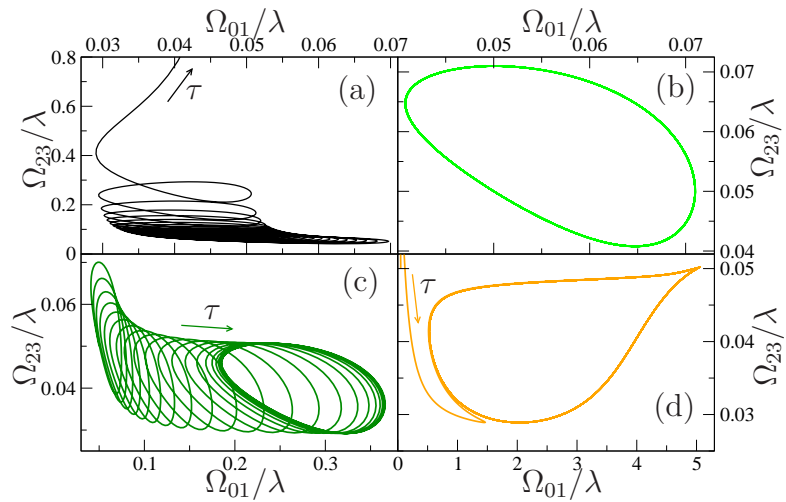}
  \caption{
     Parametric plots of coupling functions $\Omega_{01}(\tau)$ versus $\Omega_{23}(\tau)$ over the time range shown in \fig{FIG:pops_with_repump}.
     The four panels show results for different values of the amplitude recycling strength $\Omega_{03}$.
     (a) For $\Omega_{03}= 0$, the trace is unbounded as $\Omega_{23}(\tau)$ diverges.
     (b) For $\Omega_{03} = 0.1412\lambda$ the motion is purely periodic.
     (c) Above this value, the behaviour shows initial transients until once again a periodic trace is reached.
     Results are shown for $\Omega_{03} = 0.05\lambda$.
     (d) Finally, for strong amplitude recycling, $\Omega_{03}=0.5\lambda$, the driving field $\Omega_{01}(\tau)$ becomes very strongly asymmetric in time.
     Parameters as in \fig{FIG:eg_no_repump}.
     \label{FIG:JJs_with_repump}
   }
 \end{center}
\end{figure}

In \fig{FIG:JJs_with_repump} we explore the dynamics of the coupling functions and plot $\Omega_{01}(\tau)$ as a function of $\Omega_{23}(\tau)$.
Without amplitude recycling (\fig{FIG:JJs_with_repump}a) the trace simply diverges.
With $\Omega_{03}$ equal to its critical value (\fig{FIG:JJs_with_repump}b) the functions form a single closed orbit which illustrates the periodic nature of this solution.
Then, for higher values of $\Omega_{03}$ (\fig{FIG:JJs_with_repump}c), after some initial transients, the behaviour collapses onto a closed orbit.
We note that these closed orbits do not represent limit cycles because different initial values of the couplings give rise to different oscillating states.
Finally, in \fig{FIG:JJs_with_repump}d, we show results for a value of $\Omega_{03}$ far in excess of its critical value.
In this case, the long-time orbit develops a sharp kink feature, which is a sign of the pronounced lack of time-reversal symmetry in the coupling functions in this regime.

%%%%%%%%%%%%%%%%%%%%%%%%%%%%%%%%%%%%%%%%%%%%%%%
%%%%%%%%%%%%%%%%%%%%%%%%%%%%%%%%%%%%%%%%%%%%%%%

\begin{figure}[tb]
  \begin{center}
  \includegraphics[clip=true]{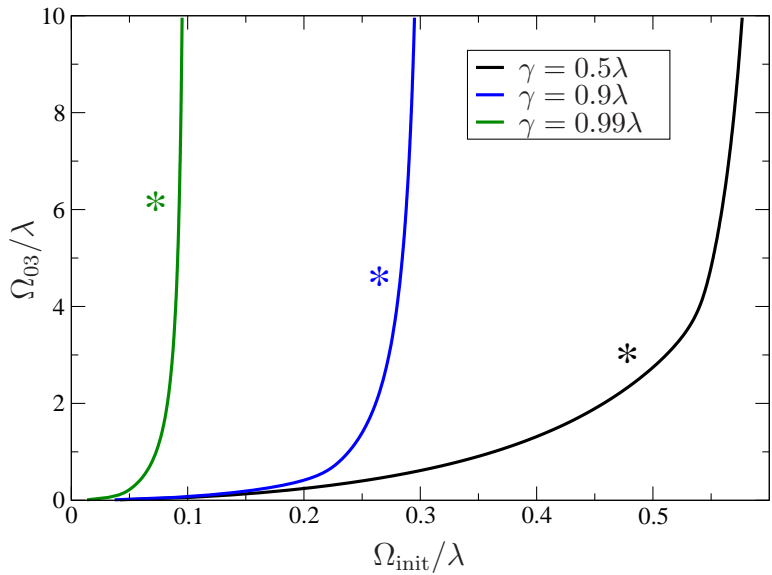}
  \caption{
     Boundaries between oscillating solutions (above the curves, marked with an asterisk) and terminating solutions (below the curves) as a function of initial coupling $\Omega_\mathrm{init}$ and amplitude-recycling strength $\Omega_{03}$ for $\gamma/\lambda=0.5,0.9,0.99$.
     Other parameters as \fig{FIG:eg_no_repump}.
     \label{FIG:boundarylines}
  }
\end{center}
\end{figure}

We have shown how the addition of amplitude recycling can elicit one of two distinct responses.
Below some critical coupling value, the emulation still terminates, albeit now with a larger breakdown time cf. the case without amplitude recycling.
Above the critical coupling, however, the amplitude-recycling coupling field stabilises the dynamics, such that the emulation enters a purely periodic mode that can run indefinitely; a situation that only occurs when $\gamma=0$ in the absence of amplitude recycling.
A clear boundary between the two regimes exists and is shown in \fig{FIG:boundarylines} as a function of $\Omega_\mathrm{init}$ and $\Omega_{03}$.

%%%%%%%%%%%%%%%%%%%%%%%%%%%%%%%%%%%%%%%%%%%%%%%
%%%%%%%%%%%%%%%%%%%%%%%%%%%%%%%%%%%%%%%%%%%%%%%

%%%%%%%%%%%%%%%%%%%%%%%%%%%%%%%%%%%%%%%%%%%%%%%%%%%%%%%%%%%%%%%%%%%%%%%%%
%%%%%%%%%%%%%%%%%%%%%      Constraints    %%%%%%%%%%%%%%%%%%%%%%%%%%%%%%%
%%%%%%%%%%%%%%%%%%%%%%%%%%%%%%%%%%%%%%%%%%%%%%%%%%%%%%%%%%%%%%%%%%%%%%%%%
\section{Experimental implementation and constraints \label{SEC:param}}

We can readily implement our amplitude recycling scheme to coherently emulate  a two-level \PT-symmetric Hamiltonian using ultracold atoms.
Let us consider a \Rb Bose-Einstein condensate in its electronic groundstate $5^2$S$_{1/2}$.
\Rb is an alkali atom with nuclear spin $I=3/2$ so that the groundstate splits into two hyperfine levels, $F=1,2$.
We identify the \PT-symmetric subspace with the $\ket{F,m_F} = \ket{1,-1}$ and $\ket{1,0}$ levels and the reservoir levels with $\ket{1,1}$ and $\ket{2,0}$ levels.
The $\ket{2,0}$ level can be connected to the stretched levels of the $F=1$ manifold via two microwave transitions that serve as the coupling to one of the reservoirs and the amplitude-recycling field.
However, binary collisions between the two hyperfine manifolds might limit the lifetime of the atomic cloud to 100\,ms.
For realistic rf coupling parameters this is in the ``long emulation time'' limit, and does not degrade the utility of this approach for simulating \PT-symmetric Hamiltonians.
Alternatively, the continuous dynamical decoupling techniques described in~\cite{trypogeorgos_synthetic_2018,anderson_continuously_2018} allow this to be equally well realized all in the $F=2$ hyperfine manifold.

Ideally, we would like to be able to emulate the \PT-symmetric system with arbitrary parameters for as long a time as desired.
While this is possible in principle, both with and without amplitude recycling, there exist a number of aspects inherent to this scheme that provide constraints on what is possible in practice.

The appearance of the timescale $\tau_\mathrm{sharp} \ll 1$, discussed in Sec.~\ref{SEC:PT}, represents a challenge since for $\gamma \to \lambda$ the period of the population oscillations diverges, whilst $\tau_\mathrm{sharp} $ remains fixed.
This timescale is directly reflected in the detunings.
Expanding \eq{EQ:E0} about one of its maxima, we obtain
$
  \delta_0(\tau) \approx
  (\lambda + \gamma)
  \left[
   1 - \gamma (\lambda + \gamma)(\tau-\tau_\mathrm{min})^2/2
  \right]
$,
which exhibits a timescale $ \left[\gamma(\gamma+\lambda)/2 \right]^{-1/2} \sim 2/\gamma = \sqrt{2}\tau_\mathrm{sharp} $.
The corollary is that experimental control must be able to simultaneously cover both small, $\sqrt{2}/\gamma$, and large, unity, time scales.

Imperfect preparation of the initial state has a similar effect in the faithful reproduction of the \PT-symmetric dynamics.
Only the squared modulus of the wavefunction, i.e., the level populations, needs to be taken into account when preparing the initial states.
The required phases can be absorbed into the driving fields through an appropriate gauge transformation, which means that control parameters do not need to be recalculated for different initial phases.
An initial error of a few percent in state preparation of one of the \PT-symmetric levels induces a small discrepancy between the realised population dynamics and the target.
Due to the larger absolute population of source and sink levels, an error of a few percent in their preparation will cause larger subsequent errors in the dynamics.
This discrepancy, however, can be made smaller if the preparation error is common-mode, where any offset in initial population is shared by both source and sink levels.
In any case, we can eliminate a state preparation error with an initial weak measurement right after state preparation takes place.
The weak measurement does not affect the initial populations, but provides the necessary information for recalibrating the dynamics of the emulation.
This way, the time-dependence of the coupling fields and detunings is dictated by the measured initial state but the subsequent dynamics always follow that of a \PT-symmetric Hamiltonian albeit with slightly different parameters than the intended ones.
Similarly, weak measurements performed while the dynamics take place can also be used to correct for errors during the emulation.

%%%%%%%%%%%%%%%%%%%%%%%%%%%%%%%%%%%%%%%%%%%%%%%
\begin{figure}[tb]
  \begin{center}
  \includegraphics[clip=true]{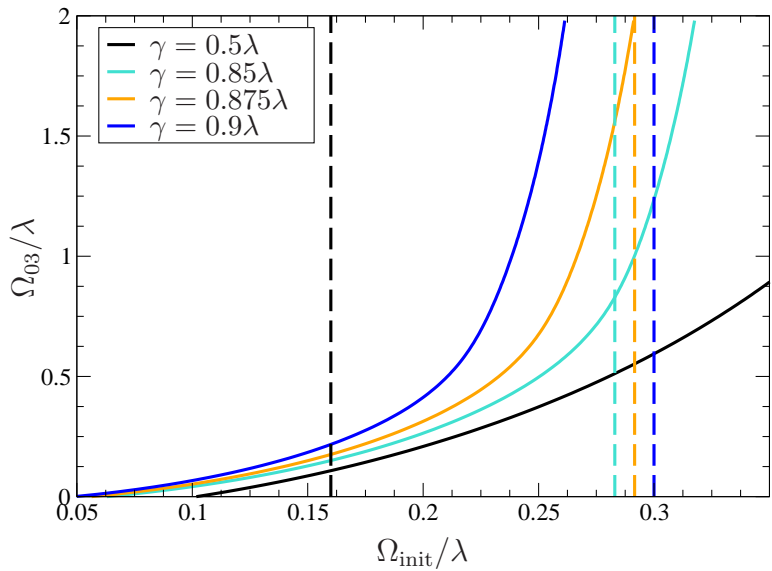}
  \caption{
     The solid lines indicate the parameter values for which we obtain a simulation break-down time  $ \tau^* = 10\pi$ for the indicated values of $\gamma$. In the regions above these curves, emulations last long enough to capture a minimum of ten Rabi oscillations of the \PT system.
     On the other hand the dashed lines indicate parameters for which $r = r_\mathrm{min}=1/10$, which from \eq{EQ:rmin} are the straight lines $\Omega_\mathrm{init} = \gamma/3$.
     Parameters to the right of these lines lead to values of the \PT-fraction $r>r_\mathrm{min}$.
     To obtain feasible emulations then, we require that these two lines cross for a given $\gamma$.
     For the parameters considered here (same as \fig{FIG:eg_no_repump}), this is the case for $\gamma/\lambda=0.5,0.85,0.875$, but not for $\gamma/\lambda=0.9$.
     \label{FIG:bound2}
  }
\end{center}
\end{figure}
%%%%%%%%%%%%%%%%%%%%%%%%%%%%%%%%%%%%%%%%%%%%%%%

Further constraints arise from the emulation scheme.
The main practical constraint limiting the total emulation time without amplitude recycling is the fact that longer times require larger populations of the source and sink levels, and this reduces the \PT-symmetric fraction $r$.
Small values of $r$ demand a large dynamic range from the experimental detection scheme.
In practice, the dynamic range of the detectors sets the minimum signal-to-noise ratio necessary for the chosen emulation parameters.
This situation however can be improved dramatically by amplitude recycling.

%%%%%%%%%%%%%%%%%%%%%%%%%%%%%%%%%%%%%%%%%%%%%%%
\begin{figure}[th]
  \begin{center}
  \includegraphics[clip=true]{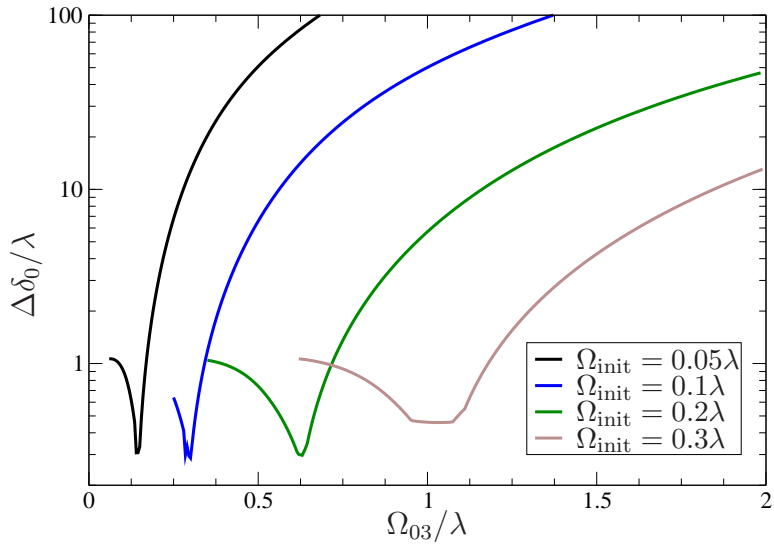}
   \caption{
     The size of the detuning range $\Delta\delta_0 \equiv |\mathrm{max}(\delta_0) - \mathrm{min}(\delta_0) |$ as a function of $\Omega_{03}$ for various values of initial couplings $\Omega_\mathrm{init}$.
     Results are only plotted here for values of  $\Omega_{03}$ that gives oscillatory solutions.
     The non-monotonic behaviour of these curves arises because the detuning $\delta_0$ has multiple extrema which change order.
     Parameters as in \fig{FIG:eg_no_repump}.
     \label{FIG:linecutE0}
   }
 \end{center}
\end{figure}
%%%%%%%%%%%%%%%%%%%%%%%%%%%%%%%%%%%%%%%%%%%%%%%

The equations in \eq{EQ:EOM} used to determine the initial wave function components of the \PT-symmetric subspace at time $\tau=0$, are independent of the amplitude recycling strength $\Omega_{03}$.
$N_0$ and therefore $r$ are unaltered by the addition of the amplitude-recycling field and \eq{EQ:rmin} applies irrespectively of $\Omega_{03}$.
Thus, the emulation times can be extended by the amplitude recycling without changing the \PT-symmetric fraction.
The minimum value of $r=r_\mathrm{min}$, set by experimental constraints, gives a minimum value for the ratio:
\begin{equation}
  \frac{\Omega_\mathrm{init}}{\gamma} \ge  \sqrt{\frac{r_\mathrm{min}}{1-r_\mathrm{min}}}
  \label{EQ:Ominitconstraint}
  ,
\end{equation}
through \eq{EQ:rmin} and this applies independently of $\Omega_{03}$.

Figure~\ref{FIG:bound2} illustrates how this constraint can be taken into account to determine which values of $\Gamma$ are experimentally accessible for a given $r_\mathrm{min}$.
Assuming we desire to simulate for a fixed time of $10\pi\alpha^{-1}$, say, we require a breakdown time $\tau^*> 10 \pi$.
This requirement defines an accessible region in the parameter space of $\Omega_\mathrm{init}$ and $\Omega_{03}$; the solid lines in the \fig{FIG:bound2} delineate the boundary of this region for various values of $\gamma$.
Let us also assume a minimum \PT-symmetric fraction of $r_\mathrm{min}=1/10$.
The constraint \eq{EQ:Ominitconstraint} then also defines a region in the same parameter space, and these boundaries are plotted with dashed lines.
Clearly, for an emulation to both run long enough and have a sufficiently large $r$, we require that the two regions overlap.
Whether this happens or not for a given $\gamma$ can be seen from \fig{FIG:bound2} as, for a feasible emulation, the two boundary lines will cross.
This happens for $\gamma/\lambda=0.5$ for example, but not for $\gamma/\lambda=0.9$.
Indeed, from this plot we can obtain the maximum value of $\gamma$ that it is possible to emulate.
If we restricted ourselves to $\Omega_{03}/\lambda \le 2$ such that the parameter range shown in \fig{FIG:bound2} is all that is accessible, then we should be able to emulate the range  $0\le \gamma /\lambda \le 0.875$.
By way of contrast, we can use the analytical approximation of \eq{EQ:tsing} to obtain the accessible range without amplitude recycling, and, for the same parameters, this turns out to be  $0 \le \gamma/\lambda \le 0.257$.
Use of the amplitude recycling therefore significantly enlarges the range of possible $\gamma$ values that can be explored.

One further constraint comes from considering the change in size of the dynamic couplings and detunings.
This is not such a problem for the couplings $\Omega_{01}$ and $\Omega_{23}$ as these are made small by construction to extend the simulation times.
On the other hand, it is desirable for the detunings to also remain small but there is no a priori reason why this should be the case.
Without amplitude recycling, \eq{EQ:E0} shows that the detunings are always bounded as
$\
  -\lambda -\gamma
  \le \delta_{0,3}(\tau)
  \le -\lambda +\gamma
  \label{EQ:E0rangenopump}
$.
These bounds change however, when we turn on the amplitude recycling. \fig{FIG:linecutE0} shows the size of the range of the detuning $\Delta \delta_0 \equiv |\mathrm{max}(\delta_0) - \mathrm{min}(\delta_0) |$ as a function of amplitude recycling strength $\Omega_{03}$.
The overall trend is that the range of values that the detunings take increases as we increase the amplitude-recycling strength.
The same quantity for detuning $\delta_3$ shows little variation and is always order $2\gamma$.
Thus, to avoid requiring large detunings, we need to operate with as small $\Omega_{03}$ as possible, in which case we obtain $\Delta\delta_0 \sim 2\gamma$.
We note that although we show the range of $\delta_0$ here, its increase is largely due to increase in magnitude of the minimum value, $|\mathrm{min}\,\delta_0|$.

%%%%%%%%%%%%%%%%%%%%%%%%%%%%%%%%%%%%%%%%%%%%%%%%%%%%%%%%%%%%%%%%%%%%%%%%%
%%%%%%%%%%%%%%%%%%%%%     DISCUSSION      %%%%%%%%%%%%%%%%%%%%%%%%%%%%%%%
%%%%%%%%%%%%%%%%%%%%%%%%%%%%%%%%%%%%%%%%%%%%%%%%%%%%%%%%%%%%%%%%%%%%%%%%%
\section{Conclusions and outlook}

Coherent emulation of \PT-symmetric or other non-Hermitian Hamiltonians can be used for gaining an understanding of these systems.
In our scheme, a single unitary system is partitioned into two subspaces coupled together with coherent couplings.
Even though the couplings are coherent, we can make them resemble particle gain and loss by appropriate manipulation of the dynamical equations describing them.
The subspace of interest is therefore made to mimic the dynamical evolution of an open system coupled to two reservoirs.
The coherent couplings between the embedded system and the `environment' can be controlled with standard experimental techniques~\cite{sugawa_observation_2016}.
This way, the effective decoherence and dynamic timescales can be adjusted arbitrarily to suit the experimental constraints.
Importantly, the choice of emulation parameters also dictates the properties of the reservoirs themselves.

Our results show how a closed, unitary system with no intrinsic gain or loss can emulate the dynamics of \PT-symmetric Hamiltonians.
In the future it will be interesting to extend these results to a wider class of non-Hermitian Hamiltonians using arbitrarily connected spin Hamiltonians as emulators.

\section*{Acknowledgements}
We are thankful to P. Solano for sparking our interest in the subject, and S. Deffner for useful discussions.
This work was initiated by DT's visit to Newcastle, funded by a Junior Research Fellowship of the Joint Quantum Centre (JQC) Durham-Newcastle.
It was partially supported by the AFOSR’s Quantum Matter MURI, NIST, and the NSF through the PFC at the JQI.

%%%%%%%%%%%%%%%%%%%%%%%%%%%%%%%%%%%%%%%%%%%%%%%%%%%%%%%%%%%%%%%%%%%%%%%%%
%%%%%%%%%%%%%%%%%%%%%      REFERENCES     %%%%%%%%%%%%%%%%%%%%%%%%%%%%%%%
%%%%%%%%%%%%%%%%%%%%%%%%%%%%%%%%%%%%%%%%%%%%%%%%%%%%%%%%%%%%%%%%%%%%%%%%%
\section*{References}
\bibliographystyle{iopart-num}
\bibliography{PTbib_no_url.bib,PT_symmetry_no_url.bib}
\end{document}